\documentstyle[12pt]{article}
\def\al{\alpha}
\def\be{\beta}
\def\ga{\gamma}

\def\ze{\zeta}

\def\ka{\kappa}

\def\rh{\rho}

\def\si{\sigma}

\def\ph{\phi}

\def\De{\Delta}
\def\Th{\Theta}

\def\fr#1#2{{{#1} \over {#2}}}
\def\prt{\partial}
\def\ap{\al^\prime}

\def\half{{\textstyle{1\over 2}}}
\def\frac#1#2{{\textstyle{{#1}\over {#2}}}}

\def\lsim{\mathrel{\rlap{\lower4pt\hbox{\hskip1pt$\sim$}}
    \raise1pt\hbox{$<$}}}
\def\gsim{\mathrel{\rlap{\lower4pt\hbox{\hskip1pt$\sim$}}
    \raise1pt\hbox{$>$}}}
\def\sqr#1#2{{\vcenter{\vbox{\hrule height.#2pt
         \hbox{\vrule width.#2pt height#1pt \kern#1pt
         \vrule width.#2pt}
         \hrule height.#2pt}}}}
\def\square{\mathchoice\sqr66\sqr66\sqr{2.1}3\sqr{1.5}3}

\newcommand{\beq}{\begin{equation}}
\newcommand{\eeq}{\end{equation}}
\newcommand{\bea}{\begin{eqnarray}}
\newcommand{\eea}{\end{eqnarray}}
\newcommand{\rf}[1]{(\ref{#1})}

\renewenvironment{thebibliography}[1]
 { \rm
   \begin{list}{\arabic{enumi}.}
    {\usecounter{enumi} \setlength{\parsep}{0pt}
     \setlength{\itemsep}{3pt} \settowidth{\labelwidth}{#1.}
     \sloppy
    }}{\end{list}}

\begin{document}
\titlepage

\begin{flushright}
{DAMTP-R93-16\\}
{IUHET 261\\}
{hep-th/9311005\\}
{September 1993\\}
\end{flushright}
\vglue 1cm

\begin{center}
{{\bf TACHYON HAIR FOR TWO-DIMENSIONAL BLACK HOLES
\\}
\vglue 1.0cm
{V. Alan Kosteleck\'y$^a$ and Malcolm J. Perry$^b$\\}
\bigskip
{\it $^a$Physics Department\\}
\medskip
{\it Indiana University\\}
\medskip
{\it Bloomington, IN 47405, U.S.A.\\}
\vglue 0.3cm
\bigskip
{\it $^b$D.A.M.T.P.\\}
\medskip
{\it University of Cambridge\\}
\medskip
{\it Silver Street\\}
\medskip
{\it Cambridge, CB3 9EW, England.\\}

\vglue 0.8cm
}
\vglue 0.3cm

\end{center}

{\rightskip=3pc\leftskip=3pc\noindent
Using a combination of analytical and numerical methods,
we obtain a two-dimensional spacetime describing
a black hole with tachyon hair.
The physical ADM mass of the black hole is finite.
The presence of tachyon hair increases the Hawking temperature.

}

\vfill
\newpage

\baselineskip=20pt

The fundamental role of black holes in the physics of the
four-dimensional real world is currently not well understood.
Two basic problems must be solved in order to elucidate
matters.
First,
we need to construct a satisfactory quantum theory of
gravitation.
Second,
we need to be able to resolve the Hawking paradox,
namely,
the apparent conflict between the laws of quantum
mechanics and the way in which black holes evaporate.

One of the ingredients in these puzzles is the observation that
black holes have a large entropy.
For non-rotating,
uncharged black holes in four dimensions,
the entropy $S$ is proportional
to the area of the event horizon
and is given by
\cite{b,h}
\beq
S = \fr {4\pi G M^2}\hbar
\quad ,
\label{a}
\eeq
where the mass of the black hole is $M$.
In contrast,
as $M$ gets large,
the entropy of most isolated systems,
e.g., stars or relativistic gas,
grows at most like $M$.
Thus, a black hole has much more entropy than a
corresponding `normal' system.
According to the Boltzmann interpretation of entropy,
this reflects a rapid growth
in the corresponding density of states $\rh (M)$.
In fact,
since
\beq
\rh(M) = e^S
\quad ,
\label{b}
\eeq
it follows that $\rh(M)$ fails to be bounded
by the Boltzmann factor $e^{-M/T}$
for large $M$
for \it any \rm value of the temperature $T$.
The negative specific heat
of the Schwarzschild black hole
is associated with this fact.

In the classical limit $\hbar \to 0$
the entropy becomes infinite,
which is reflected in the black-hole uniqueness theorems
(the `no-hair' theorems).
This powerful collection of results in classical general
relativity shows that the only degrees of freedom of black
holes observable from outside the event horizon are
those related to longitudinal degrees of freedom of gauge
fields that are also physical.
Thus,
a black hole is characterized by its mass,
momentum, and angular momentum
(from the gravitational field),
and electric and magnetic charge
(from a $U(1)$ gauge field)
\cite{b2,c,bu,m}.
Similar considerations apply to non-abelian Yang-Mills fields
and systems with Higgs fields
\cite{p,br,bk,sz,gmo}.

One testing ground
for exploration of these puzzles
is the case of two spacetime dimensions,
where a satisfactory nonperturbative version
of string theory exists,
thereby providing a quantum theory of gravity.
Within this framework,
a black-hole solution has recently been found
\cite{msw,w,gp1,f}.
The entropy of this black hole,
at least in the sigma-model approximation,
is given by
\beq
S = M/T_c
\quad .
\label{c}
\eeq
Here,
$M$ is the ADM mass
\cite{adm}
and $T_c$ is the temperature,
which is a fixed parameter
derived from the central charge of the theory.

This two-dimensional result is dramatically different from
that found in four dimensions.
The entropy only grows linearly with the mass,
like normal systems.
Consequently,
the density of states is bounded by the Boltzmann factor
provided the physical temperature $T$ obeys $T<T_c$.
Presumably there is a phase transition at $T=T_c$,
but the physical significance of this is not currently understood.
In any event,
the slower growth of the density of states in two dimensions
suggests that the black hole uniqueness theorems might
not have as broad an applicability in $d=2$ as in $d=4$.
This would appear to be especially important for the phase
$T<T_c$.

Let us examine the uniqueness theorems at the semiclassical
level for the case of two-dimensional bosonic string theory.
The physical states of the theory are divided into two classes:
the tachyon field $T$,
which is a massless scalar field,
and an infinite set of discrete states,
of which the graviton and dilaton are examples.
In the $\si$-model approximation to string theory,
target-space physics can be described by an effective
spacetime action
\cite{cfmp}
\beq
I = - \fr 1 {2 \ka^2} \int d^2x \sqrt g e^\ph\left( \hat c - R -
(\nabla\ph)^2 + (\nabla T)^2 + 2V(T) \right)
\quad ,
\label{d}
\eeq
where
$R$ is the Ricci scalar of the metric $g_{ab}$,
$\phi$ is the dilaton,
and $T$ is the tachyon with potential
$V(T)= -2T^2/\ap + \ldots$.
The constant $\hat c$ is related to the central charge of the string
theory and is given by
\beq
\hat c = \fr {2(d-26)}{3\ap}
\quad ,
\label{e}
\eeq
while $\ap$ is the inverse string tension
and $\ka$ corresponds to the Newton constant in two dimensions.

Varying the action
\rf{d}
with respect to the metric
and tachyon yields the corresponding beta-function equations
at one loop:
\beq
R_{\mu\nu} = \nabla_\mu\nabla_\nu \phi + \nabla_\mu T\ \nabla_\nu T
\quad
\label{f}
\eeq
for the metric,
and
\beq
\square\ T + \nabla_\mu\phi ~ \nabla^\mu T
= \fr{\prt V}{\prt T}
\quad
\label{g}
\eeq
for the tachyon.
Variation of the action with respect to the dilaton
gives an equation consistent with the Bianchi identities,
representing a first integral of Eq.\ \rf{g}
with $\hat c$ as the constant of integration.
It can be written
\beq
\hat c = R-(\nabla\phi)^2
- 2~\square\ \phi -(\nabla T)^2 - 2V(T)
\quad .
\label{h}
\eeq

One expects black-hole equilibrium states to be static
and asymptotically flat,
so we assume the spacetime prosesses
a timelike Killing vector $k^\mu$,
at least in the asymptotic region.
Since the black hole is an isolated object,
at large distances from it
the target space can be taken as the linear dilaton vacuum.
A general form of the metric for a static $d=2$ spacetime is
\beq
ds^2 = -v^2 dt^2 + \fr{dr^2}{w^2}
\quad ,
\label{i}
\eeq
where $v$ and $w$ are taken as functions of $r$ only.
With this parametrization,
the beta-function equation
\rf{f} for the metric becomes the two equations
\beq
\fr{v^{\prime\prime}}v + \fr{v^\prime}v \fr{w^\prime}w
+\fr{v^\prime}v \ph^\prime = 0
\quad
\label{jz}
\eeq
and
\beq
\fr{v^{\prime\prime}}v + \fr{v^\prime}v \fr{w^\prime}w
+ \ph^{\prime\prime} +\fr{w^\prime}w \ph^\prime
+ T^{\prime 2} = 0
\quad ,
\label{kz}
\eeq
where a prime denotes $d/dr$.
The tachyon equation \rf{g} becomes
\beq
T^{\prime\prime}
+\left(\fr{v^\prime}v + \fr{w^\prime}w + \ph^\prime\right) T^\prime
- \fr 1 {w^2} \fr{\prt V}{\prt T}
= 0
\quad ,
\label{lz}
\eeq
and the dilaton equation \rf{h} becomes
\beq
\fr{v^{\prime\prime}}v + \fr{v^\prime}v \fr{w^\prime}w
+ \ph^{\prime\prime} +\half\ph^{\prime 2}
+\left(\fr{v^\prime}v + \fr{w^\prime}w\right) \ph^\prime
+ \half T^{\prime 2}
+ \fr 1 {w^2} \left( V + \half \hat c\right) = 0
\quad .
\label{mz}
\eeq

To simplify formulae in what follows,
we adopt units such that $\ap = 2$
and we keep only the quadratic terms in the tachyon potential.
This means, for example, that
$V(T) = -T^2$ and $\hat c = -8$.
Integrating Eq.\ \rf{j} and rearranging the other equations
yields the convenient forms
\beq
v^{\prime}= \fr k w e^{-\ph}
\quad ,
\label{j}
\eeq
where $k$ is an integration constant,
\beq
\ph^{\prime\prime}
+\left(\fr{v^\prime}v + \fr{w^\prime}w + \ph^\prime\right) \ph^\prime
+ \fr 2 {w^2}(4+T^2) = 0
\quad ,
\label{k}
\eeq
\beq
T^{\prime\prime}
+\left(\fr{v^\prime}v + \fr{w^\prime}w + \ph^\prime\right) T^\prime
+ \fr 2 {w^2} T = 0
\quad ,
\label{l}
\eeq
and the first integral,
\beq
T^{\prime 2}=
\ph^{\prime 2} + 2\fr{v^\prime}v \ph^{\prime}
- \fr 2 {w^2} (4+T^2 )
\quad .
\label{m}
\eeq
These equations contain the essential information
about the metric-dilaton-tachyon system.

If the tachyon background vanishes, $T=0$,
then there are only two relevant solutions to these
equations.
Their expression is simplified
by working with the convenient local-gauge choice
$w=1$.
The first solution is the linear-dilaton vacuum,
in which the spacetime is flat and the dilaton is given by
\beq
\ph = \ph_0 + \sqrt{8}~ r
\quad .
\label{n}
\eeq
The second is the black-hole solution
for which
\beq
v(r) = \tanh \sqrt{2}~r
\quad
\label{o}
\eeq
and
\beq
\ph(r) = \ph_0 + 2\ln\cosh \sqrt{2}~r
\quad .
\label{p}
\eeq
As $r\to\infty$,
the black-hole spacetime tends to the linear-dilaton vacuum.
The mass of the black hole is given by
\cite{w,gp1,f}
\beq
M = \sqrt{8} e^{\ph_0}
\quad .
\label{q}
\eeq
The precise definition of $M$ is discussed below.
The black hole has an event horizon at $r = 0$,
where $v(r) = 0$.
Assuming that $v \to 1$ at spatial infinity,
the Hawking temperature $T_H$
of the horizon is given generally by
\cite{gp2}
\beq
T_H = \fr 1{2\pi} v^\prime\vert_{\rm horizon}
\quad
\label{r}
\eeq
and so,
evaluating it in this case,
we find
\beq
T_H = \fr 1{\sqrt{2}~\pi} =T_c
\quad .
\label{s}
\eeq

An immediate question about tachyonic hair
is whether one can prove a no-hair theorem
analogous to those for general relativity.
A no-hair theorem for scalar fields
in two dimensions,
valid under certain conditions,
has been presented in
Ref.\ \cite{nl}.
However,
this result does not apply here because the
tachyon has derivative couplings to
the dilaton.

Given the absence of a definitive result excluding
tachyon hair,
the first step is to examine Eqs.\ \rf{j} -- \rf{m}
to see if we can find tachyon hair
for the case of linearized perturbations about the
black-hole metric.
Thus,
taking Eq.\ \rf{l} and using Eqs.\ \rf{o} and \rf{p}
for the background,
we find
\beq
T^{\prime\prime}
+ \sqrt{2}
\left( \fr
{1+2 \sinh ^2\sqrt{2}~r}
{\sinh \sqrt{2}~r ~~\cosh \sqrt{2}~r}
\right) T^\prime + 2 \sqrt 2 ~T = 0
\quad .
\label{t}
\eeq
This is essentially a hypergeometric equation
\cite{dvv}.
Define the new radial coordinate
\beq
\xi = \cosh^2 \sqrt{2}~r
\quad .
\label{u}
\eeq
This maps the event horizon $r=0$ into $\xi =1$
and $r$-spatial infinity into $\xi$-spatial infinity.
The spacetime singularity,
which is not covered by the coordinate system
defined by Eq.\ \rf{o},
now appears at $\xi = 0$.
Equation \rf{t} becomes
\beq
8 \xi (\xi - 1)\fr {d^2 T}{d\xi^2}
+ 8(2\xi - 1) \fr {dT}{d\xi}
+ 2 T = 0
\quad .
\label{v}
\eeq
This hypergeometric equation is represented by
\beq
\cal{P} \left\{
\matrix{0&1&\infty&~~\cr
        0&0&\half&;~\xi\cr
        0&0&\half&~~\cr}
\right\}
\quad
\label{w}
\eeq
in the Riemann-Papperitz scheme.
Near $\xi = 1$, the two linearly independent solutions
are of the form
\beq
T \sim {\rm constant}~~,~~~~
T \sim \ln \vert\xi - 1\vert
\quad ,
\label{x}
\eeq
while as $\xi \rightarrow \infty$ they are of the form
\beq
T \sim \fr 1 {\sqrt{\xi}}~~,~~~~
T \sim \fr {\ln \vert\xi \vert}{\sqrt{\xi}}~
\quad .
\label{y}
\eeq
The solution that is regular at the horizon is given by
\beq
T
=\fr{2T_0}{\pi} {\rm sech} \sqrt 2~ r ~K(\tanh\sqrt 2 ~r)
= T_0~ P_{-1/2}(2\xi - 1)
= T_0~{_2F_1}(\half,\half;1;1-\xi)
\quad ,
\label{z}
\eeq
where
$K$ is the complete elliptic integral of the first kind,
$T_0$ is the value of the tachyon field at the horizon,
$P_{-1/2}$ is the Legendre function of
order $-\half$,
sometimes called a conical function,
and
${_2F_1}$ is the hypergeometric function.
A plethora of details about the latter can be found in
Ref.\ \cite{s}.
The first representation is convenient for
examining the behavior of $T$ at spatial infinity,
whereas the latter two are convenient for
the behavior near the horizon.
The other solution to Eq.\ \rf{w} diverges at the
horizon and therefore is not physically acceptable.

Whilst regularity of the fields at infinity and at the horizon
are necessary conditions for a physically acceptable
solution,
we must also check that the energy in the field is finite.
The energy-momentum tensor $\Th_{\mu\nu}$ of the tachyon
field is given by
\beq
\Th_{\mu\nu} = \prt_\mu T~\prt_\nu T
- \half g_{\mu\nu}~ g^{\rh\si}~\prt_\rh T ~\prt_\si T
\quad .
\label{aa}
\eeq
Therefore,
the energy in the field in the domain of outer communication
is
\beq
E_T = \int_{\cal{H}}^\infty k^\mu k^\nu \Th_{\mu\nu}\ dr
\quad ,
\label{ba}
\eeq
where $\cal{H}$ is the event horizon.
Substituting for the specific form
of the metric \rf{o} and the tachyon perturbation \rf{z} leads
to the expression
\beq
E_T = \half \int_{0}^\infty \tanh^2 \sqrt{2}~r~ T^{\prime 2}~ dr
\quad .
\label{ca}
\eeq

Amusingly, Eq.\ \rf{ca} can be written in terms of
a generalized hypergeometric function.
Since
\beq
T = T_0~{_2F_1}(\half,\half;1;-\sinh^2 \sqrt{2}~r)
\quad ,
\label{da}
\eeq
it follows that
\beq
T^\prime = \fr {-T_0} {\sqrt{2}}
\sinh\sqrt 2 ~ r ~\cosh\sqrt 2 ~ r ~
{_2F_1}(\frac 32,\frac 32;2;-\sinh^2 \sqrt{2}~r)
\quad .
\label{ea}
\eeq
By using the quadratic transformations of Gauss and Kummer,
we can rewrite this expression as
\beq
T^\prime = \fr {-T_0} {\sqrt{8}}
\sinh \sqrt 8 ~ r ~
{_2F_1}(\frac 34,\frac 34;2;-\sinh^2\sqrt{8}~r)
\quad .
\label{fa}
\eeq
Clausen has developed an expression for the square of
this function
\cite{cl},
which gives
\beq
T^{\prime 2} = \fr {T_0^2} 8
\sinh^2 \sqrt 8 ~ r ~
{_3F_2}(\frac 32,\frac 32,\frac 32;2,3;-\sinh^2 \sqrt{8}~r)
\quad .
\label{ga}
\eeq
Thus,
\beq
E_T = \frac 1 4 T_0^2 \int_{0}^\infty dr ~
\sinh^4 \sqrt{2}~ r ~~
{_3F_2}(\frac 32,\frac 32,\frac 32;2,3;-\sinh^2 \sqrt{8}~r)
\quad .
\label{ha}
\eeq

Although an analytical form for this integral might exist,
we do not pursue this approach further here.
Instead,
we restrict ourselves to proving that this expression
is finite.
Both functions ${_2F_1}$ and ${_3F_2}$ have singularities
only when their argument takes the values $0$, $1$,
or $\infty$.
This follows because they obey second- and third-order
differential equations,
respectively,
that have ordinary points
everywhere except for regular singular points
at these three locations.
Expanding $T(r)$ for small $r$ gives
\beq
T(r) = T_0(1 - \half r^2 + \frac {11}{48}r^4 + \ldots )
\quad .
\label{ia}
\eeq
Thus, for small $r$
the integrand in Eq.\ \rf{ha}
is of order $r^6$, and so the contribution to the integral
is finite.
As $r$ increases to infinity,
the argument of the hypergeometric function decreases,
and so the only remaining potential singularity in the integral
appears as $r \to \infty$.
However,
in this limit
\beq
T(r) \longrightarrow \fr{4T_0}{\pi} e^{-\sqrt 2 ~r}
\left( \sqrt 2 ~r + \ln 2 + {\rm O}(\fr 1 r) \right)
\quad .
\label{ja}
\eeq
The integrand in Eq.\ \rf{ha}
therefore varies as $e^{-\sqrt 8 ~r}$
in the limit as
$r \to \infty$,
so finiteness is guaranteed.
The finiteness of the perturbative contribution
to the energy strongly suggests that there is enough
energy in the tachyon field to further curve the
spacetime,
but insufficient energy to destroy asymptotic flatness.

Our analysis so far makes plausible
the existence of black-hole spacetimes
with tachyon hair.
To prove that this indeed occurs,
the complete nonlinear equations
\rf{j} -- \rf{m} must be solved,
and the solutions must be shown to have
the required finiteness properties.
This is sometimes termed the `back-reaction'
problem.
It has been examined by several authors
in the context of various approximation schemes.
Ref.\ \cite{al} considers lowest-order perturbation theory
and argues that the event horizon remains regular.
In contrast,
Ref.\ \cite{ra} sets $V(T) = 0$ and
claims that the event horizon becomes singular.
Ref.\ \cite{mo}
uses perturbation theory
to argue that the energy of the spacetime is infinite,
a conclusion also reached in
Ref.\ \cite{pst}.
Finally,
Ref. \ \cite{cm} argues that no black-hole spacetime
is possible for nonzero tachyon background.
The situation is evidently rather confused.

One practical approach to settling this issue
is the numerical integration of
Eqs.\ \rf{j} -- \rf{m}.
It suffices to integrate the three equations
\rf{j}, \rf{k}, and \rf{m}.
For the numerical analysis,
it is convenient to work in the the gauge $w=1$
and to convert these three equations
into four first-order equations.
We have performed the numerical integration
using a fourth-order generalized Runge-Kutta method
implemented on an HP Apollo Series 700 workstation.
The program is initialized as follows.
Values of the tachyon charge $T_0$
and the dilaton strength $\ph_0$
at the horizon $r=0$ are selected.
As described below,
perturbation theory in $r$ near the horizon,
applied for the full nonlinear equations,
is then used to fix the initial values of
$v$, $\ph$, $\ph^\prime$ and $T$ at some specified
small $r$.
The integration is continued until the asymptotic regimes
of the fields are attained.

Figures 1-4 show the results for a tachyon charge
$T_0 = 0.1$ with $\ph_0 = 0$.
The associated metric is displayed in Figure 1.
The invariance of Eqs.\ \rf{j}--\rf{m}
under scaling of the metric $v$ has been used
to set the asymptotic value to the canonical choice of one.
As can be seen,
the metric is a smooth function of $r$ that rises
to close to its asymptotic value within about two
radial units.
Figure 2 displays the dilaton,
while Figure 3 shows its slope.
The dilaton begins at zero and within about two radial units
converts to a linearly rising trajectory,
with asymptotic slope $\sqrt 8 \simeq 2.8$.
In contrast,
as can be seen in Figure 4,
the tachyon begins at $T_0 = 0.1$
and falls smoothly,
approaching zero after about six radial units.
The results indicate that solutions to the nonlinear
equations do indeed exist
and that $v$, $\ph$ and $T$ are smooth functions everywhere
outside the event horizon.

It is useful to examine the solutions analytically near
the horizon and asymptotically at spatial infinity
to gain insight into the physics.
There are two natural methods
for extracting results for the event horizon:
using perturbation theory either in $r$ or in $T_0$.
Both yield insights,
so we present them in turn.

First, consider the expansion in positive powers of $r$.
The symmetries of Eqs.\ \rf{j} -- \rf{m}
in the gauge $w=1$ imply that the tachyon and dilaton
can be expanded in even powers of $r$,
while the metric can be expanded in odd powers.
Substitution into Eqs.\ \rf{j} -- \rf{m}
and collection of coefficients
generates three series for the
metric, the dilaton, and the tachyon.
The coefficients can be written as functions of the
square of the tachyon charge, $T_0^2$.
We use the lower-order terms in these three series
to initialize our numerical analysis.

For the tachyon,
we find
\beq
T=T_0 \left(1 - \half r^2
+ \fr{11+2T_0^2}{48} r^4 - \ldots \right)
\quad .
\label{jaa}
\eeq
This provides the nonlinear correction at order $T_0^2$
to Eq.\ \rf{ia},
which was obtained in the linearized theory.
For the dilaton,
we get the equation
\beq
\ph=\ph_0  + \half(4+T_0^2)r^2 - \ldots
\quad .
\label{jab}
\eeq
The constant $\ph_0$ can be set to zero without loss of
generality,
as it merely represents a shift in the definition
of the origin of the dilaton field.
Finally, for the metric
we obtain
\beq
v=v_1 \left(r - \fr{4+T_0^2}{6} r^3 + \ldots\right)
\quad .
\label{jac}
\eeq
Here,
$v_1$ corresponds to the arbitrary scale choice that
can be made for the metric.
We have adopted the convention that $v_1$ is fixed
so that $v\to 1$ as $r\to\infty$.
Using Eq.\ \rf{r},
it then follows that
\beq
v_1=2 \pi T_H
\quad .
\label{jad}
\eeq

Further results for the event horizon can be extracted
via an expansion in powers of $T_0$
about the analytical black-hole solution \rf{o} and \rf{p}.
Taking Eq.\ \rf{z} for the tachyon field,
we write
\beq
v=\tanh\sqrt 2 ~r + T_0^2 g(r) + {\rm O}(T_0^4)
\quad ,
\label{ka}
\eeq
The dilaton can be eliminated from Eq.\ \rf{m} using
Eq.\ \rf{j}.
In the gauge $w=1$,
Eq.\ \rf{m} then reduces to a second-order differential
equation for the metric perturbation $g$,
given by
\bea
- 2T^2 - T^{\prime 2}
&=&
2\left({\rm coth}\sqrt2 ~r
+ 2\sinh\sqrt 2 ~r ~\cosh\sqrt 2 ~r
\right) g^{\prime\prime}
\nonumber \\
&+& 8 \sqrt 2 \sinh^2 \sqrt 2~r ~g^\prime
+ \fr 8 {\sinh\sqrt 2 ~r ~~\cosh\sqrt 2 ~r} g
\quad .
\label{laa}
\eea
The relevant solution of this equation satisfies
the boundary conditions $g\to v_1 r$ as $r\to 0$
and $g\to 0$ as $r\to\infty$.
By treating the left-hand side of this equation
as a source and obtaining the Green function for
the second-order differential operator on the
right-hand side,
an integral expression for $g$ satisfying the
boundary conditions can be found.

The explicit form is not needed here.
Instead,
we seek the value of $g^\prime (0)$,
since this information gives the Hawking temperature
as a function of $T_0^2$
via Eqs.\ \rf{jad} and \rf{ka},
\beq
T_H(T_0) = \fr 1 {2\pi} \left (
\sqrt 2+g^\prime (0)T_0^2 + {\rm O}(T_0^4) \right)
\quad .
\label{lab}
\eeq
The Green-function method yields the integral expression
\beq
g^\prime (0)=
\frac 1 4 \int_0^\infty ~dr~
\fr{\sinh \sqrt 8 ~r}{\cosh^2\sqrt 8 ~r}
\left( 2 T^2 + T^{\prime 2}\right)
\quad .
\label{lac}
\eeq
The integrand is positive.
This implies that the Hawking temperature
is increased above $T_c$ when a tachyon charge is added
to the black hole.
Qualitatively,
this effect is different from the situation for the
Reissner-Nordstrom solutions in four dimensions,
where the addition of electric charge \it decreases \rm
the Hawking temperature instead.

Although the integral might be performed analytically
using the explicit expressions for the tachyon
field given in Eq.\ \rf{z},
we do not pursue this here.
In practice our numerical methods suffice to determine
the Hawking temperature for any given situation.
For example,
we find numerically that
$T_H \simeq 0.2253$
for the solution displayed
in Figures 1 to 4.
In contrast,
the uncharged black hole has
$T_H \simeq 0.2251$.
In fact,
we have determined that
the Hawking temperature of the black hole
with tachyon hair is given approximately as
\beq
T_H=\fr 1 {\sqrt 2 ~ \pi} \left( 1 + \frac 1 {10} T_0^2
+ {\rm O}(T_0^4) \right)
\quad .
\label{nab}
\eeq

In principle,
an analytical expression for $T_H$
might be found that is exact to all orders in $T_0$.
However,
in practice this is difficult
because Eqs.\ \rf{r} and \rf{jad} only
hold if $v \to 1$ at spatial infinity,
so knowledge of the asymptotic behavior of $v$, $\ph$, and $T$
associated with a particular behavior near the horizon
is required.

Next, we turn to an investigation
of the analytical behavior of our solutions
at infinity.
The gauge choice $w=1$
is less convenient for this issue,
so instead we define a new radial coordinate $\ze$
such that
\beq
\ph = \tilde\ph_0 + \ln \ze
\quad ,
\label{oa}
\eeq
where $\tilde \ph_0$ is a constant.
In the absence of a tachyon background, $T=0$,
$\ph_0 = \tilde\phi_0$,
and the new coordinate $\ze$ reduces to the coordinate $\xi$
previously introduced.
In this gauge,
Eq.\ \rf{k} becomes
\beq
\dot T^2 = \fr 1 \ze\left(\fr {\ddot v}{\dot v}
+ \fr {\dot v} v\right) +\fr 2 {\ze^2}
\quad ,
\label{pa}
\eeq
where the dot signifies a $\ze$ derivative.

Given a functional form for $T$,
this equation can be integrated directly to give $v$.
As $r\to\infty$,
the asymptotic form of $T$ is given by Eq.\ \rf{ja}.
Note that this equation is only valid for small $T_0$ because there
are corrections to the tachyon equation of order $T_0^2$
coming from the metric and dilaton.
Since $\ze = \cosh^2 \sqrt 2~r$,
the tachyon can be written as
\cite{bf}
\beq
T=\fr {T_0} \pi \fr 1 {\sqrt\ze}
\left( \ln\ze + 4 \ln 2
\right)
+ {\rm O}\left(\ze^{-\frac 3 2}\right)
\quad .
\label{qa}
\eeq

Integrating Eq.\ \rf{pa} yields
\beq
\ze^2 v \dot v =
C\exp\left(\int\ze \dot T^2 ~d\ze\right)
\quad ,
\label{ra}
\eeq
where $C$ is a integration constant.
Substituting for $\dot T$ using Eq.\ \rf{qa}
gives
\beq
v\dot v =
\fr C{\ze^2} \exp\left[\fr 1 {4 \pi^2 \ze} T_0^2
\left( - (\ln\ze)^2 + (2 - 8 \ln2)\ln\ze
- 2 + 8 \ln 2 - 16(\ln 2)^2
\right)
+ {\rm O}(\ze^{-2})
\right].
\label{sa}
\eeq
As $\ze \to \infty$ we can
approximate this expression by taking only the first
terms in the expansion of the exponential.
Integrating the resulting expression gives
\beq
\half v^2 = D - \fr C \ze +
\fr C {8 \pi^2 \ze^2} T_0^2
\left( (\ln\ze)^2 - \ln\ze(1 - 8 \ln2)
+\frac 3 2 - 4 \ln 2 + 16(\ln 2)^2
\right)
+ {\rm O}(\ze^{-3}),
\label{taz}
\eeq
where $D$ is another integration constant
that sets the scale of the time coordinate.
We choose the conventional time coordinate
with $v\to 1$ as $\ze\to\infty$,
which gives $D=\half$.
Thus,
\bea
v=1 - \fr C \ze
&+& \fr C {8 \pi^2 \ze^2} T_0^2
\Biggl( (\ln\ze)^2 - \ln\ze(1 - 8 \ln2)
\nonumber \\
&&\quad\quad\quad\quad
+\frac 3 2 - 4 \ln 2 + 16(\ln 2)^2 -\fr{2\pi^2 C}{T_0^2}
\Biggr)
+ {\rm O}(\ze^{-3})
\quad .
\label{ta}
\eea
This analytical expression correctly reproduces the
behavior of our numerical solutions.

With these results,
the metric coefficient $w$ in the new gauge
can be found from Eq.\ \rf{j}.
Asymptotic flatness together with the conventional
normalization requires that
$w\to\sqrt 8~\ze$ as $\ze \to \infty$.
This fixes the value of $k$ in terms of $C$ and $\tilde\ph_0$:
\beq
k=\sqrt 8 ~C~e^{\tilde\ph_0}
\quad .
\label{va}
\eeq
Explicit evaluation of $w$ leads to
\bea
w= \sqrt 8~ \ze
&-& \fr 1 {\sqrt 2~ \pi^2} T_0^2
\Biggl( -(\ln\ze)^2 + \ln\ze(2 - 8 \ln2)
\nonumber \\
&&\quad\quad\quad\quad
- 2 + 8 \ln 2 - 16(\ln 2)^2 +\fr{2\pi^2 C}{T_0^2}
\Biggr)
+ {\rm O}(\ze^{-1})
\quad .
\label{wa}
\eea

The remaining issue is the physical meaning of the
constant of integration $C$.
Rescaling the coordinate $\ze$
allows the elimination of $C$ from the system of
equations.
Therefore,
$C=1$ is a legitimate choice.
However,
in the process the gauge condition \rf{oa}
must be preserved.
The effect of this rescaling is therefore to
renormalize the value of $\tilde\ph_0$.
Henceforth,
we choose $C=1$.

Next,
we calculate the ADM mass of this spacetime.
The canonical treatment
\cite{dw}
proceeds as follows.
Starting from the action \rf{d},
one first needs to obtain the hamiltonian.
This requires a separation of the coordinates
into space and time,
denoted by $x$ and $t$ here.
In the remainder of this paper,
we redefine a prime to indicate an $x$ derivative
and a dot to indicate a $t$ derivative.

The metric decomposes in the usual manner as
\beq
g_{\mu\nu} = \left(\matrix{-\al^2+\be^2/\ga&\be\cr
			   \be&\ga\cr}\right)
\quad .
\label{vaz}
\eeq
The variable $\ga$ can be regarded as a one-dimensional
spatial metric.
The next step is substitution of this metric form
into the action and
the determination of the canonical momenta conjugate
to the canonical coordinates
$\al$, $\be$, $\ga$, $\ph$, and $T$.
These are
\beq
\pi_\al = 0~~,~~~~\pi_\be = 0~~,~~~~
\pi_\ga = \fr 1{\al\sqrt\ga} \dot \ph e^\ph
\quad
\label{waz}
\eeq
and
\beq
\pi_\ph = \fr{2\sqrt \ga}{\al}\dot T e^\ph
{}~~,~~~~
\pi_T = -\fr{2\sqrt \ga}{\al}\dot \ph e^\ph -
\fr 1{\al\sqrt\ga} \dot \ga e^\ph
\quad .
\label{xa}
\eeq
Both $\pi_\al$ and $\pi_\be$ vanish by virtue of
the constraints of the system.
For simplicity,
we have worked in the gauge $\be = 0$.

For time-independent field configurations,
the canonical hamiltonian then becomes
\beq
H= \int dx~e^\ph\left(
-\al\sqrt\ga \hat c
- \fr {2\al^\prime}{\sqrt\ga}\ph^\prime
- \fr {\al}{\sqrt\ga}\ph^{\prime 2}
+ \fr {\al}{\sqrt\ga}T^{\prime 2}
-\al\sqrt\ga~ V(T) \right)
\quad .
\label{ya}
\eeq
However,
in a reparametrization-invariant system,
$H$ must vanish modulo a boundary term.
This boundary term is the ADM energy of the
system.
To convert $H$ into a boundary term,
it suffices to substitute the beta-function
equation for the dilaton,
Eq.\ \rf{mz},
into Eq.\ \rf{ya}.
This gives
\beq
H= \int dx~\left[- \fr 2{\sqrt\ga} e^\ph\left(
\al^\prime
+\al\ph^\prime\right)\right]^\prime
\quad .
\label{za}
\eeq
Thus, the ADM energy is
\beq
E=  \fr 2 {\sqrt\ga} e^\ph\left(
\al^\prime
+\al\ph^\prime\right)\biggr\vert_{x\to\infty}
\quad .
\label{ab}
\eeq

In two dimensions,
$E$ is infinite even for flat space.
The physically important quantity for a curved
spacetime $\cal{M}$ with given metric and dilaton
on the boundary at infinity is the
difference between the ADM energy $E_{\cal{M}}$
and the corresponding quantity $E_{\cal{F}}$
for flat spacetime $\cal{F}$
with the \it same \rm  values of the metric and dilaton
on the boundary.

For our black hole with tachyon charge,
the first step in obtaining the ADM mass
is to construct a flat metric
such that it coincides at $\ze=\hat\ze$, say,
with that of the tachyonic black hole.
The canonically normalized linear-dilaton
vacuum,
which in the present variables is
given by Eq.\ \rf{oa}
and
\beq
v=\al = 1~~,~~~~w=\fr 1{\sqrt \ga} = \sqrt 8~\ze
\quad ,
\label{bb}
\eeq
does not satisfy this requirement.
Instead,
suppose that at $\ze = \hat \ze$
the variables $v$, $w$, $\ph$
take the values $\hat v$, $\hat w$, $\hat\ph$,
respectively.
Then,
flat space has the form
\beq
\ph = \hat \ph_0 + \ln \fr \ze{\hat\ze}~~,~~~~
v=\al = \hat v~~,~~~~w=\fr 1{\sqrt \ga} = \sqrt 8~\ze
\quad .
\label{cb}
\eeq
This can be compared with the metric of the
tachyonic black hole,
Eqs.\ \rf{ta} and \rf{wa},
at the point $\hat\ze$.
The fiducial flat-space metric has the same
values of $v$, $w$, and $\ph$ at that point.

Subtracting the ADM energy of the flat
spacetime $\cal{F}$ from that of the curved
spacetime $\cal{M}$ and cancelling terms where possible
gives
\beq
\De E=E_{\cal{F}}- E_{\cal{M}}
=2 e^{\hat\ph}\hat w\fr{\prt v}{\prt\ze}
\quad .
\label{db}
\eeq
Explicitly,
\beq
\De E=2e^{\tilde\ph_0}\ze
\left(\sqrt 8~\ze +{\rm O}\left( (\ln\ze)^2\right)\right)
\left( \fr 1 {\ze^2} + {\rm O}\left(\fr 1 {\ze^3}\right)\right)
\quad .
\label{eb}
\eeq
For large $\ze$,
this expression has a well-defined and finite limit,
which we interpret as the physical mass $M$ of the
tachyonic black hole.
We find
\beq
M=\sqrt {32}~e^{\tilde\ph_0}
\quad .
\label{fb}
\eeq
Note that
this general expression agrees
with the results for the uncharged black hole
\cite{w,gp1,f},
up to an overall numerical factor
that arises because the constant $\tilde\ph_0$
has not been canonically fixed.

We thus conclude that in two spacetime dimensions
it is possible to find black holes with arbitrary
finite physical mass and tachyon charge,
with Hawking temperature above that of the
corresponding uncharged black holes.

\vglue 0.4cm

We thank the Aspen Center for Physics for hospitality.
This work was supported in part
by the North Atlantic Treaty Organization
under grant number CRG 910192,
by the United States Department of Energy
under grant number DE-FG02-91ER40661,
and by Trinity College, Cambridge.

\vglue 0.6cm
{\bf\noindent Figure Captions}
\vglue 0.4cm

Figure 1. The metric $v$ as a function of $r$
in dimensionless units.

Figure 2. The dilaton $\ph$ as a function of $r$
in dimensionless units.

Figure 3. The dilaton slope $\phi^\prime$ as a function of $r$
in dimensionless units.

Figure 4. The tachyon $T$ as a function of $r$
in dimensionless units.


\begin{thebibliography}{xx}

\bibitem{b}
J.D. Bekenstein,
Lett. Nuov. Cim. 4 (1972) 737;
Phys. Rev. D 7 (1973) 2333.

\bibitem{h}
S.W. Hawking, Commun. Math. Phys. 43 (1975) 199.

\bibitem{b2}
J.D. Bekenstein,
Phys. Rev. D 5 (1972) 1239, 2403.

\bibitem{c}
B. Carter,
in S.W. Hawking and W. Israel, eds.,
\it General Relativity: An Einstein
Centenary Survey \rm
(Cambridge University Press, Cambridge, 1979).

\bibitem{bu}
G. Bunting,
Ph.D. thesis, Univ. of New England, Armidale,
New South Wales, 1983.

\bibitem{m}
P. Mazur,
in M.A.H. MacCallum, ed.,
\it General Relativity and Gravitation \rm
(Cambridge University Press, Cambridge, 1987).

\bibitem{p}
M.J. Perry,
Phys. Lett. 71B (1977) 234.

\bibitem{br}
F.A. Bais and R.J. Russell,
Phys. Rev. D 11 (1975) 2692.

\bibitem{bk}
R. Bartnik and J. McKinnon,
Phys. Rev. Lett. 61 (1988) 141.

\bibitem{sz}
N. Straumann and Z.-H. Zhou,
Nucl. Phys. B360 (1991) 180;
Phys. Lett. B 237 (1990) 353.

\bibitem{gmo}
B.R. Greene, S.D. Mathur, and C.M. O'Neill,
Phys. Rev. D 47 (1993) 2242.

\bibitem{msw}
G. Mandal, A.M. Sengupta, and S. Wadia,
Mod. Phys. Lett. A6 (1991) 1685.

\bibitem{w}
E. Witten, Phys. Rev. D 44 (1991) 314.

\bibitem{gp1}
G.W. Gibbons and M.J. Perry,
Int. J. Mod. Phys. D1 (1992) 335.

\bibitem{f}
V. Frolov, Phys. Rev. D 46 (1992) 5383.

\bibitem{adm}
R. Arnowitt, S. Deser, and C.W. Misner,
in L. Witten, ed.,
\it Gravitation: an Introduction to Current Research \rm
(Wiley, New York, 1962).

\bibitem{cfmp}
C. Callan, D. Friedan, E. Martinec and M.J. Perry,
Nucl. Phys. B262 (1985) 593.

\bibitem{gp2}
G.W. Gibbons and M.J. Perry,
Proc. Roy. Soc. A358 (1978) 467.

\bibitem{nl}
C. Nappi and O. Lechtenfeld,
Phys. Lett. B 288 (1992) 72.

\bibitem{dvv}
R. Dijkgraaf, H. Verlinde, and E. Verlinde,
Nucl. Phys. B371 (1992) 269.

\bibitem{s}
L. Slater,
\it Generalized Hypergeometric Functions \rm
(Cambridge University Press, Cambridge, 1966).

\bibitem{cl}
T. Clausen,
J. Reine Angew. Math. 3 (1828) 89.

\bibitem{al}
S.P. de Alwis and J. Lykken,
Phys. Lett. B269 (1991) 264.

\bibitem{ra}
S.K. Rama,
Dublin preprints hepth 9302075 and 9303140 (1993).

\bibitem{mo}
N. Marcus and Y. Oz,
Tel Aviv preprint hepth 9305003 (1993).

\bibitem{pst}
A. Peet, L. Susskind and L. Thorlacius,
Stanford preprint hepth 9305030 (1993).

\bibitem{cm}
S. Chaudhuri and D. Minic,
Austin preprint hepth 9301105 (1993).

\bibitem{bf}
P.F. Byrd and M.D. Friedman,
\it Handbook of Elliptic Integrals \rm
(Springer Verlag, Berlin, 1954).

\bibitem{dw}
B.S. DeWitt,
Phys. Rev. 160 (1967) 1113.

\end{thebibliography}
\end{document}